\documentclass[comsoc]{IEEEtran}
\usepackage{cite}
\usepackage{amsmath,amssymb,amsfonts}
\usepackage{algorithmic}
\usepackage{graphicx}
\usepackage{textcomp}
\usepackage{gensymb}
\usepackage{stfloats}
\usepackage{soul}
\usepackage{subcaption}
\usepackage{balance}

\usepackage{xcolor}
\def\BibTeX{{\rm B\kern-.05em{\sc i\kern-.025em b}\kern-.08em
    T\kern-.1667em\lower.7ex\hbox{E}\kern-.125emX}}

\begin{document}

\title{Multi-Hop Quantum Key Distribution with Passive Relays over Underwater Turbulence Channels\\
}

\author{\IEEEauthorblockN{Amir Hossein Fahim Raouf\IEEEauthorrefmark{1}, Majid Safari\IEEEauthorrefmark{2} and Murat Uysal\IEEEauthorrefmark{1}}\\
\IEEEauthorblockA{\IEEEauthorrefmark{1}Dept. of Electrical and Electronics Engineering, Ozyegin University, Istanbul, Turkey\\
}
\IEEEauthorblockA{\IEEEauthorrefmark{2}Institute for Digital Communications, University of Edinburgh,
Edinburgh, UK}
}


\maketitle
\addtolength{\topmargin}{+0.29cm}
\begin{abstract}
Absorption, scattering, and turbulence experienced in underwater channels severely limit the range of quantum communications. In this paper, to overcome range limitations, we investigate a multi-hop underwater quantum key distribution (QKD) where intermediate nodes help the key distribution between the source and destination nodes. We consider deployment of passive-relays which simply redirect the qubits to the next relay node or receiver without any measurement. Based on near-field analysis, we present the performance of relay-assisted QKD scheme in clear ocean under different atmospheric conditions. We further investigate the effect of system parameters (aperture size and detector field-of-view) on the achievable distance.
\end{abstract}


\section{Introduction}
Today's cryptosystems such as widely deployed RSA and elliptic curve-based schemes build upon the formulation of some intractable computational problems. They are able to offer only computational security within the limitations of conventional computing power. Recent advances in the quantum computing towards the so-called quantum supremacy have the potential to eventually break such classical cryptosytems \cite{1, 2}. Based on the firm laws of quantum mechanics rather than some unproven foundations of mathematical complexity, quantum cryptography provides a radically different solution for key distribution promising unconditional security \cite{3}. 

The current literature on quantum key distribution (QKD) mainly focuses on fiber optic, atmospheric and satellite links \cite{4}. The increasing deployment of underwater sensor networks (USNs) particularly for harbour and maritime surveillance and protection as well as the need for secure underwater communication for military needs (e.g., submarine communication) have further prompted researchers to investigate underwater QKD \cite{5, 6, 7, 8, 9, 10, 11, 12, 13, 14}. In particular, the quantum bit error rate (QBER) and secret key rate of well-known BB84 protocol were studied in \cite{8, 14}. The performance of other QKD protocols such as entanglement \cite{13} and decoy state \cite{6} were further investigated in underwater environments. In addition to these theoretical and simulation studies, experimental works were also conducted to demonstrate the feasibility of underwater QKD \cite{7, 10, 11, 12}.

The above experimental and theoretical studies point out that performance degradation due to absorption, scattering, and turbulence experienced in underwater channels severely limit the range of quantum communication links. In this paper, to overcome range limitations, we investigate relay-assisted underwater QKD where intermediate nodes help the key distribution between the source and destination nodes. While the concept of relay-assisted QKD was earlier studied for atmospheric, fiber and satellite links \cite{15, 16, 17}, it was not yet studied, to the best of our knowledge, for underwater quantum links. 

In this paper, we consider a multi-hop underwater QKD system where relay nodes are utilized along the path connecting two legitimate parties. Unlike classical optical communication systems \cite{18}, amplify-and-forward and detect-and-forward relaying cannot be used in QKD since any type of measurement modifies the quantum state \cite{3}. To address this, we utilize passive relays \cite{17} which simply redirect the qubits to the next relay node or to the destination node without performing any measurement or detection process. Such relays can be implemented by adaptive optics \cite{19, 20, 21} to reconstruct the turbulence-degraded wave-front of the received beam and redirect the resulting collimated beam. Under the assumption of passive relays and based on a near-field analysis \cite{22} over underwater turbulence channels, we derive an upper bound on QBER. Based on this upper bound, we present the performance of underwater QKD in different atmospheric conditions (i.e., clear weather with full moon at night and heavy overcast when sun is near horizon) and different levels of turbulence strength. We further investigate the effect of detector field-of-view (FOV) and aperture size on the system performance. 

The remainder of this paper is organized as follows. In Section \ref{sys_model}, we describe our relay-assisted system model based on BB84 QKD protocol. In Section \ref{Per_anlz}, we derive an upper bound on the QBER of the system under consideration over underwater turbulence channels. In Section \ref{sim_res}, we present numerical results and finally, we conclude in Section \ref{concd}.

\section{System Model} \label{sys_model}
\begin{figure}[tb]
\centering
\includegraphics[width=0.95\linewidth]{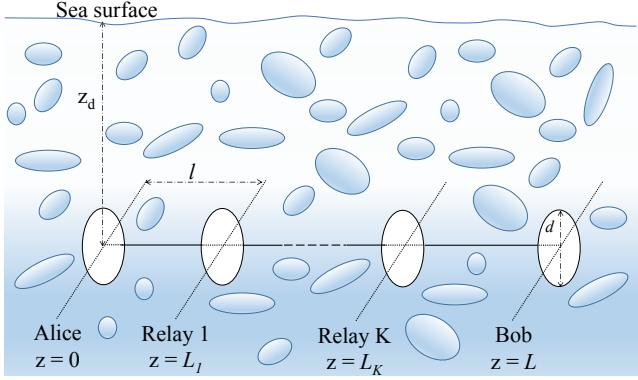}
\caption{Underwater BB84 QKD system under consideration.}
\label{fig:1}
\end{figure}

We consider a relay-assisted underwater QKD system with $K$ serial passive relay nodes over a link distance of $L$. As illustrated in Fig. \ref{fig:1}, Alice (transmitter) with a diameter size of $d$ is placed in $z = 0$ plane. Relay nodes and Bob (receiver) with the same diameter size of $d$ are located in the $z = L_i$. The consecutive nodes in the serial scheme are placed equidistant along the path from the source to the destination. Therefore, the length of each hop is equal to $l = {L \mathord{\left/
 {\vphantom {L {\left( {K + 1} \right)}}} \right.
 \kern-\nulldelimiterspace} {\left( {K + 1} \right)}}$.

The QKD system is built upon BB84 protocol \cite{23} which aims to create a secret key between the authorized partners (Alice and Bob) such that eavesdropper (Eve) fails to acquire meaningful information. BB84 protocol is based on the principle of polarization encoding. In this protocol, Alice prepares a qubit by choosing randomly between two linear polarization bases namely rectilinear (denoted by $\oplus$) or diagonal (denoted by $\otimes$) for every bit she wants to send. She selects a random bit value ``0" or ``1" and uses polarization encoding of photons where polarization of ${{0^\circ } \mathord{\left/
 {\vphantom {{0^\circ } { - 45^\circ }}} \right.
 \kern-\nulldelimiterspace} { - 45^\circ }}$ represents 0 and polarization of ${{ + 90^\circ } \mathord{\left/
 {\vphantom {{ + 90^\circ } { + 45^\circ }}} \right.
 \kern-\nulldelimiterspace} { + 45^\circ }}$ represents 1.

At the receiver side, Bob measures the arriving photon randomly in either $\oplus$ or $\otimes$ bases. Alice and Bob determine the secure key with respect to the received qubits at the ``sift" events. Sift events occurs at the bit intervals in which
exactly one of the single photon
detectors registers a count and both Alice and Bob have chosen the same basis.
 Alice and Bob can recognize the sift events by transferring information over a public classical communication channel (in our case underwater optical channel). Based on the sifted qubits, a shared one-time pad key is created to use for secure communication \cite{24}. 

Alice generates each qubit with an average photon number of ${n_S}$ which is encoded with the corresponding polarization state of the qubit for a randomly chosen basis. As a result of underwater path loss and turbulence, the $i^{th}$ relay ($i = 1, \ldots ,K$) collects only a random fraction ${\gamma_i}$ of the transmitted photons. Under the assumption of identical transmitter/receiver sizes and equidistant placement, we can simply write ${\gamma_1 = \gamma_2 = \ldots = \gamma_K = \gamma}$.  The relay node forwards the captured photons to the next relay (or Bob) by redirecting the light beam and without any measurements. Therefore, Bob will collect an overall fraction ${\gamma _{Bob}} = {\gamma ^{K + 1}}$ of the originally transmitted photons from Alice. 

In addition to the received photons from the source, receiver of each relay node will collect some background noise. The total average number of background photons per polarization at the $i^{th}$ relay can be therefore expressed as
\begin{equation}
{n_{{B_i}}} = {n_{B0}} + {n_{B0}}\gamma  + {n_{B0}}{\gamma ^2} +  \ldots  + {n_{B0}}{\gamma ^{i - 1}}
\end{equation}
The accumulated background photons at Bob's receiver therefore becomes \cite{17}
\begin{equation}
{n_B} = {n_{B0}}\frac{{1 - {\gamma ^{K + 1}}}}{{1 - \gamma }}
\end{equation}
Beside background noise, each of Bob's detectors will collect dark current noise. Let $I_{dc}$ denotes the dark current count rate. The average number of dark counts is given by ${n_D = {I_{dc}}\Delta t}$. Thus, the average number of noise photons reaching each Bob's detector can be obtained by ${n_N} = {{{n_B}} \mathord{\left/
 {\vphantom {{{n_B}} 2}} \right. \kern-\nulldelimiterspace} 2} + {n_D}$. It should be noted that since the relays just redirect the photons, they do not increase the dark current.
 
 \section{Performance Analysis}\label{Per_anlz}
 In this section, we investigate the performance of the underwater QKD system through the derivation of an upper bound on QBER. QBER is the ratio of probabilities of sift and error which depend on the statistical characteristics of received fraction of transmitted photons $\gamma$. This can be expressed as ${\gamma = h\left( l \right)\hat \mu}$ where $h\left( l \right)$ is the deterministic path loss term and $\hat \mu$ is the random channel coefficient, also called as ``power transfer" in \cite{17}. This channel coefficient defines the probability of transmitted photon being reliably received considering the channel turbulence ($0 \le \hat \mu  \le 1$) \cite{new25}.
 
 As discussed in \cite{22}, finding a statistical description of $\hat \mu$ and therefore $\gamma$ is a formidable task and a closed-form expression is not available in the literature. As an alternative, an upper bound on QBER was presented in \cite{17} for a terrestrial relay-assisted QKD system using an upper bound on the noise count and based on \textit{average power transfer} defined as ${\mu  \buildrel \Delta \over = {\mathop{\rm E}\nolimits} \left[ {\hat \mu } \right] }$.This is given by \cite[Eq. (28)]{17}, shown at the top of next page.
 \begin{figure*}[t]
\begin{equation}
 {\rm{QBER}} \le \frac{{2\eta {{\hat n}_N}\exp \left[ { - \eta 4{{\hat n}_N}} \right]\left( {1 - {\mu ^{K + 1}} + \exp \left[ { - \eta {n_S}{h^{K + 1}}\left( l \right)} \right]{\mu ^{K + 1}}} \right)}}{{b\exp \left[ { - b} \right]\left( {1 - c} \right) + \left( {a + b} \right)\exp \left[ { - \left( {a + b} \right)} \right]c}}
 \label{eq:3}
\end{equation}
 \hrule
\end{figure*}

In (\ref{eq:3}), $\eta$ is the quantum efficiency of Geiger-mode avalanche photodiodes (APDs) and ${\hat n_N}$ is an upper bound on the noise count, i.e., ${n_N} \le {\hat n_N}$, whose derivation will be elaborated later. In (\ref{eq:3}), $a$, $b$ and $c$ are defined by
 \begin{equation}
     a = \eta \left[ {{n_S}{h^{K + 1}}\left( l \right) + 2{n_{B0}}\left( {\frac{{1 - {h^{K + 1}}\left( l \right)}}{{1 - h\left( l \right)}} - 1} \right)} \right]
 \end{equation}
 \begin{equation}
     b = \eta \left( {2{n_{B0}} + 4{n_D}} \right)
 \end{equation}
 \begin{equation}
     c = \frac{{{n_S}{{\left( {\mu h\left( l \right)} \right)}^{K + 1}} + 2{n_{B0}}\left( {\frac{{1 - {{\left( {\mu h\left( l \right)} \right)}^{K + 1}}}}{{1 - \mu h\left( l \right)}} - 1} \right)}}{{{n_S}{h^{K + 1}}\left( l \right) + 2{n_{B0}}\left( {\frac{{1 - {h^{K + 1}}\left( l \right)}}{{1 - h\left( l \right)}} - 1} \right)}}
 \end{equation}
 
 The calculation of $h\left( l \right)$, $\mu$ and ${\hat n_N}$ depends on the operation environment. In the following, we elaborate on their calculations for the underwater channel under consideration. The underwater path loss is a function of attenuation and geometrical losses. For collimated light sources, the geometrical loss is negligible; therefore, the path loss with a laser diode transmitter only depends on the attenuation loss. The attenuation loss is characterized by wavelength-dependent extinction coefficient $\varsigma$. Typical value of extinction coefficient for clear ocean can be found in Table \ref{table1} for $\lambda = 532$ nm \cite{25}. In our work, we utilize the modified version of Beer-Lambert formula proposed in \cite{26}, which takes into account the contribution of scattered lights. The underwater path loss for a transmission distance of $l$ can be expressed as
 
 \begin{equation}
     h\left( l \right) = \exp \left[ { - \varsigma l{{\left( {\frac{d}{{\theta l}}} \right)}^T}} \right]
 \end{equation}
 where $\theta$ is the full-width transmitter beam divergence angle and $T$ is a correction coefficient, \cite{26}.
 
 The average power transfer for each hop (i.e., a distance of $l$ over turbulent path) can be expressed as \cite{22}
 
 \begin{equation}\label{eq:8}
     \mu \! =\!\!\frac{{8\sqrt F }}{\pi }\!\!\int\limits_0^1 \!{e^{(\frac{- W({dx,l})}{2})}
     \!\left( {{{\cos }^{ - 1}}\left(\! x \right)\! -\! x\sqrt {1 - {x^2}} }\! \right)\!{J_1}\left( {4x\sqrt F } \right)\!dx}
 \end{equation}
where $J_1 (\cdot)$ is the first-order Bessel function of the first kind and $F$ is the Fresnel number product of transmit and receive diameters given by ${F = {\left( {\pi {d^2}/4\lambda l} \right)^2}}$. In ({\ref{eq:8}}), $W (\cdot,\cdot)$ is the underwater wave structure function. For given transmission distance of $l$ and given separation distance between two points on the phase front transverse to the axis of propagation (denoted by $\rho$), it is given by \cite{14}
\begin{equation}\label{eq:9}
\begin{split}
    W\left( {\rho ,l} \right) = & 1.44\pi {k^2}l\left( {\frac{{{\alpha ^2}\chi }}{{{\omega ^2}}}} \right){\varepsilon ^{ - \frac{1}{3}}}\left( {1.175\eta _K^{2/3}\rho  + 0.419{\rho ^{\frac{5}{3}}}} \right)\times \\
    & \left( {{\omega ^2} + {d_r} - \omega \left( {{d_r} + 1} \right)} \right)
    \end{split}
\end{equation}
where $k = {{2\pi } \mathord{\left/
 {\vphantom {{2\pi } \lambda }} \right.
 \kern-\nulldelimiterspace} \lambda }$ is the wave number, $\omega$ is the relative strength of temperature and salinity fluctuations, $\varepsilon$ is the dissipation rate of turbulent kinetic energy per unit mass of fluid, $\alpha$ is the thermal expansion coefficient, $\chi$ is the dissipation rate of mean-squared temperature and ${d_r}$ is the eddy diffusivity ratio. In ({\ref{eq:9}}), ${\eta_K}$ is Kolmogorov microscale length and given by ${\eta_K} = {\left( {{{{\upsilon ^3}} \mathord{\left/
 {\vphantom {{{\upsilon ^3}} \varepsilon }} \right.
 \kern-\nulldelimiterspace} \varepsilon }} \right)^{{1 \mathord{\left/
 {\vphantom {1 4}} \right.
 \kern-\nulldelimiterspace} 4}}}$ with $\upsilon$ referring to the kinematic viscosity. 
 
 In an underwater environment, the primary source of noise is the refracted sunlight from the surface of the water. Let ${R_d}\left( {\lambda ,{z_d}} \right)$ denote the irradiance of the underwater environment as a function of wavelength and underwater depth. With respect to sea surface (i.e., ${z_d = 0}$), it can be written as ${R_d}\left( {\lambda ,{z_d}} \right) = {R_d}\left( {\lambda ,0} \right){e^{ - {K_\infty }{z_d}}}$ where ${K_\infty}$ is the asymptotic value of the spectral diffuse attenuation coefficient for spectral down-welling plane irradiance \cite{27}. The typical total irradiances at sea level, i.e., ${R_d}\left( {\lambda ,0} \right)$, in the visible wavelength band for some typical atmospheric conditions are provided in \cite{28}. The background photons per polarization on average can be then given by \cite{29}
 \begin{equation}\label{eq:10}
     {n_{B0}} = \frac{{\pi {R_d}A\Delta t'\lambda \Delta \lambda \left( {1 - \cos \left( \Omega  \right)} \right)}}{{2{h_p}{c_0}}}
 \end{equation}
where $A$ is the receiver aperture area, $\Omega$ is the FOV of the detector, ${h_p}$ is Planck's constant, $\Delta \lambda$ is the filter spectral width, $\Delta t$ is the bit period and $\Delta t'$ is the receiver gate time. Ignoring the effect of turbulence (i.e., $\hat \mu = 1$) on the redirected background photons coming from relays \cite{17}, an upper bound on the noise count at each of Bob's four detectors can be obtained as
\begin{equation}\label{eq:11}
    {\hat n_N} = \frac{{{n_{B0}}}}{2}\left( {\frac{{1 - \exp \left[ { - \varsigma {L^{1 - T}}{{\left( {\frac{{d\left( {K + 1} \right)}}{\theta }} \right)}^T}} \right]}}{{1 - \exp \left[ { - \varsigma {{\left( {\frac{L}{{K + 1}}} \right)}^{1 - T}}{{\left( {\frac{d}{\theta }} \right)}^T}} \right]}}} \right) + {n_D}
\end{equation}

\textbf{Special case 1:} As a sanity check, consider ${K = 0}$ (i.e., no relay). Therefore, ${\hat n_N}$, $a$ and $c$ can be simplified as ${\hat n_N = {{{n_{B0}}} \mathord{\left/
 {\vphantom {{{n_{B0}}} 2}} \right.
 \kern-\nulldelimiterspace} 2} + {n_D} = {b \mathord{\left/
 {\vphantom {b {4\eta }}} \right.
 \kern-\nulldelimiterspace} {4\eta }}}$, $a = \eta {n_S}h\left( L \right)$ and $c = {\mu_L}$. Replacing these in ({\ref{eq:3}}) yields
 
 \begin{equation}
{\rm{QBER}} \le \frac{{{{\hat n}_N}\left[ {1 \!- \!{\mu _L} + \!{\mu _L}{e^{ - \eta {n_S}h\left( L \right)}}} \right]}}{{\frac{{{n_S}{\mu _L}h\left( L \right)}}{2}{e^{ - \eta {n_S}h\left( L \right)}} \!+ \!2{{\hat n}_N}\!\left[ {1 \!- \!{\mu _L} + \!{\mu _L}{e^{ - \eta {n_S}h\left( L \right)}}} \right]}}
 \end{equation}
 where $h\left( L \right)$ and ${\mu _L}$ are respectively the path loss and the average power transfer for the length of line-of-sight link connecting Alice and Bob. It can be readily checked that this result coincides with \cite[Eq. (4)]{14} which was earlier reported for underwater QKD link.

\textbf{Special case 2:} 
As another benchmark, we consider a QKD system operating over non-turbulent condition. The exact QBER of such a QKD system is given by \cite{22}

\begin{equation}\label{eq:13}
    {\rm{QBE}}{{\rm{R}}_{non}} = \frac{{2{n_N}}}{{{n_S}{\mu _0}h\left( L \right) + 4{n_N}}}
\end{equation}
where ${\mu _0}$ is the largest eigenvalue of the singular value decomposition of vacuum-propagation Green’s function given in \cite{30}.

\section{Simulation Results}\label{sim_res}
In this section, we demonstrate the performance of relay-assisted underwater QKD scheme under consideration. We assume the transmitter beam divergence angle of $\theta = {6^ \circ }$, the dark current count rate of ${I_{dc}} = 60$ Hz, filter spectral width of $\Delta \lambda = 30$ nm, bit period of $\Delta t = 35$ ns, receiver gate time of $\Delta t' = 200$ ps, Geiger-Mode APD quantum efficiency of $\eta = 0.5$ and ${n_S} = 1$. Unless otherwise stated, we assume the transmitter and receiver aperture diameters of $d = 5$ cm, FOV of $\Omega = {180^ \circ }$ and clear atmospheric conditions at night with a full moon. We consider clear ocean as water type. As for channel parameters, we assume ${\alpha_{th} = 2.56\times10^{-4} \;\rm{1/deg}}$ and ${\upsilon = 1.0576\times 10^{-6} \;\rm{m^{2}s^{-1}}}$ \cite{31}. For turbulence model, we assume $\omega = -2.2$, ${\chi_T = 10^{-5} \;\rm{K^{2}s^{-3}}}$ and ${\varepsilon = 10^{-5} \;\rm{m^{2}s^{-3}}}$ which corresponds to strong oceanic turbulence \cite{32}. For the convenience of the reader, the channel and system parameters are summarized in Table \ref{table1}.

\begin{table}[ht]
\caption{System and channel parameters}
\renewcommand{\arraystretch}{1.3}
\label{table1}
\begin{center}
\scalebox{0.75}{
\begin{tabular}{ |l|l|l| } 
 \hline
 \textbf{Parameter} & \textbf{Definition} & \textbf{Numerical Value} \\ \hline
$\Omega$ & \text{Field of view} & $180\degree$ \cite{26} \\ \hline 
$\Delta \lambda$ & \text{Filter spectral width} & $30$ $\rm{nm}$ \cite{14} \\ \hline 
$\lambda$ & \text{Wavelength} & $530$ $\rm{nm}$ \cite{26} \\ \hline 
$\Delta t$ & \text{Bit period} & $35$ $\rm{ns}$ \cite{29} \\ \hline 
$\Delta t\ensuremath{'}$ & \text{Receiver gate time}  & $200$ $\rm{ps}$ \cite{29} \\ \hline 
$d$ & \text{Aperture diameter} & $5$ $\rm{cm}$ \cite{22} \\ \hline 
$\eta$ & \text{Quantum efficiency} & $0.5$ \cite{22} \\ \hline 
$I_{dc}$ & \text{Dark current count rate} & $60$ $\rm{Hz}$ \cite{29} \\ \hline 
$K_{\infty}$ & \text{Asymptotic diffuse attenuation coefficient} & $0.08$ $\rm{m^{-1}}$ \cite{27} \\ \hline 
$z_d$ & \text{Depth} & $100$ $\rm{m}$ \cite{29} \\ \hline 
$\theta$ & \text{Transmitter beam divergence angle} & $6\degree$ \cite{26} \\ \hline 
$\varsigma$ & \text{Extinction coefficient for clear water} & $0.151$ $\rm{m^{-1}}$ \cite{25} \\\hline
$T$ & \text{Correction coefficient} \begin{tabular}{c|c}  & $\theta  = 6^\circ ,\,\,{d} = 5\,{\rm{cm}}$ \,\,\,\,\,\,\,\, \\
 & $\theta = 6^\circ ,\,\,{d} = 10\,{\rm{cm}}$\,\,\,\,\,\,\,\, \\
 & $\theta = 6^\circ ,\,\,{d} = 20\,{\rm{cm}}$ \,\,\,\,\,\,\,\, \\
 & $\theta = 6^\circ ,\,\,{d} = 30\,{\rm{cm}}$ \,\,\,\,\,\,\,\,
 \end{tabular} & 
 \begin{tabular}{l}  \!\!\!\! $0.13$  \cite{26}\\
 \!\!\!\! $0.16$  \cite{26}\\
\!\!\!\! $0.21$  \cite{26} \\
\!\!\!\! $0.26$  \cite{26} \end{tabular} \\\hline
\end{tabular}}
\end{center}
\end{table}

\begin{figure}[bht]
\centering
\includegraphics[width=0.95\linewidth]{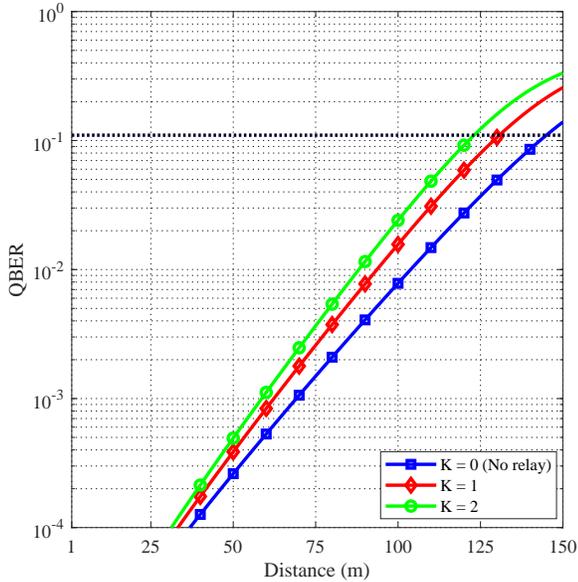}
\caption{The QBER of the relay-assisted QKD system for clear ocean with no turbulence at night time with a full moon for $K = 0, 1, 2$ relay nodes.}
\label{fig:2}
\end{figure}

Fig. \ref{fig:2} illustrates the performance of QBER with respect to the total link distance for non-turbulent clear ocean. We assume $K = 1, 2$ relay nodes. The exact QBER for direct link, i.e. $K = 0$, based on (\ref{eq:13}) is also included as a benchmark. It can be observed from Fig. \ref{fig:2} that relaying fails to improve the QBER performance in non-turbulent channel conditions. The main reason for such behavior is that adding passive relay nodes leads to additional collected background noise at the receiver. 

\begin{figure}[tb]
\centering
\includegraphics[width=0.95\linewidth]{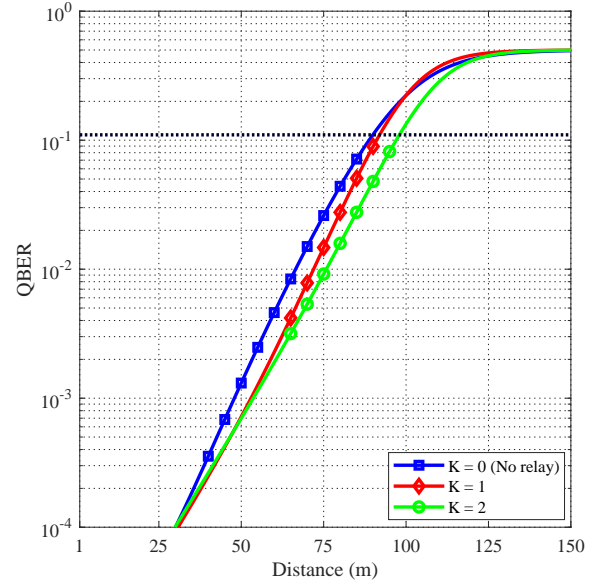}
\caption{Effect of strong turbulence in clear ocean at night time with a full moon on QBER for $K = 0$, $1$, $2$ relay nodes.}
\label{fig:3}
\end{figure}

In Fig. \ref{fig:3}, we now investigate the performance of relay-assisted QKD in the presence of turbulence. Specifically, we present the upper bound on QBER with respect to link distance based on (\ref{eq:3}) for clear ocean with strong turbulence. It can be observed that, unlike non-turbulent conditions, relaying now improves the QBER performance in the presence of turbulence. This is due to the fact that shorter hops mitigate the effect of turbulence-induced fading on the QBER performance. For instance, to achieve ${\rm{QBER}} \le 11\%$\footnote{It is generally accepted that for BB84 protocol is secure against a sophisticated quantum attack if QBER is less than 11\% \cite{33}. }, the achievable distance for direct link is $89$ m. It increases to $92$ m and $98$ m with $K = 1$ and $K = 2$ relay nodes, respectively. 

\begin{figure}[tb]
\centering
\includegraphics[width=0.95\linewidth]{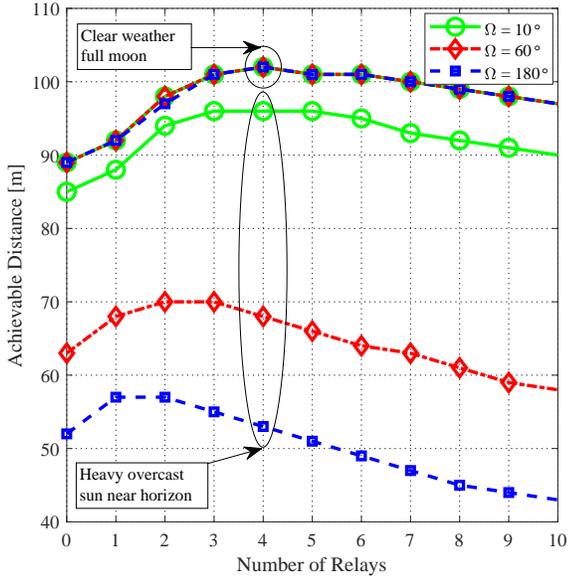}
\caption{Achievable distance versus the number of relay for different FOV values.}
\label{fig:4}
\end{figure}

In Fig. \ref{fig:4}, we study the effect of FOV on the performance of relay-assisted underwater QKD system. Specifically, we present the achievable distance versus the number of relay nodes. As atmospheric conditions, we assume clear weather with full moon and heavy overcast when sun is near horizon. For FOV values, we have $\Omega = {10^ \circ}$, ${60^ \circ}$ and ${180^ \circ}$. It is observed that at night, the achievable distances are almost identical and independent of FOV values, i.e., all three plots overlap with each other. The maximum achievable distance is obtained as $102$ m when we employ four relay nodes. However, further increase in relay nodes does not improve the performance since, according to (\ref{eq:11}), increasing the number of relay nodes leads to an increase in the background noise redirected from relays to Bob’s receiver. 

Benefit of choosing a proper value of FOV becomes clear as the environment irradiance increases. Our results demonstrate that the optimal number of relay (in the sense of maximizing the achievable distance) increases as the FOV decreases. As can be readily checked from (\ref{eq:10}), the effect of FOV on ${n_{B0}}$ is more pronounced at day time due to higher value of environment irradiance. Thus, increasing FOV results in increase of the background noise at each relay node and consequently, this increases the background noise redirected from relays to Bob’s receiver. For example, in daylight, the maximum achievable distance for $\Omega = {180^ \circ}$ is $57$ m which can be attained by employing one relay. On the other hand, the achievable distance for $\Omega = {60^ \circ}$ attains its maximum value with two relays while the maximum achievable distance for $\Omega = {10^ \circ}$ is obtained with three relay nodes.

\begin{figure}[tb]
\centering
\includegraphics[width=0.95\linewidth]{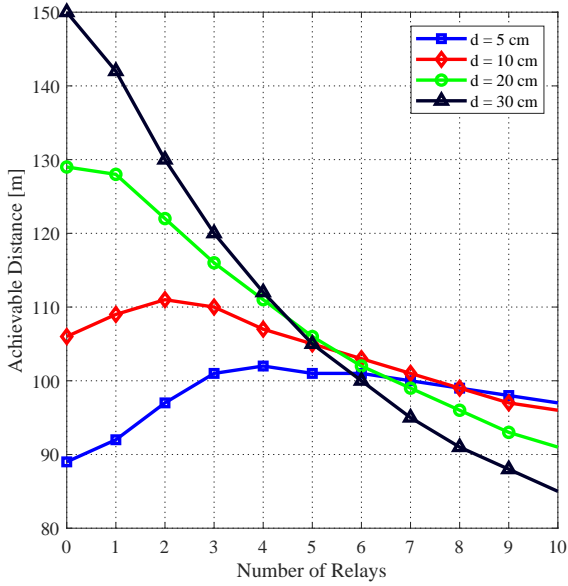}
\caption{Achievable distance versus the number of relay for different aperture sizes.}
\label{fig:5}
\end{figure}

In Fig. \ref{fig:5}, we present the achievable distance versus the number of relay nodes for different aperture sizes. We assume clear weather with full moon. For aperture sizes, we consider $d = 5$, $10$, $20$ and $30$ cm. It is observed that as the diameter decreases the optimal number of relay node increases. For instance, the optimal number of relay nodes (in the sense of maximizing the achievable distance) for $d = 5$ and $10$ cm is four and two, respectively. However, it is observed that adding relay nodes fails to improve the achievable distance for $d = 20$ and $30$ cm, i.e. the direct communication (no relay case) outperforms the relay scheme. It is due to the fact that although larger diameter result in an increase of collected photons coming from Alice, this also increases the average number of background photons at Bob’s receiver. 

\section{Conclusions}\label{concd}
In this paper, we have investigated the performance of relay assisted underwater QKD based on the BB84 protocol in the presence of turbulence. Our results have demonstrated that turbulence-induced fading can be mitigated by adding passive relay nodes which leads to an improvement in the achievable distance. Although relaying increases the average number of collected photons coming from Alice, it also results in an increase of the average number of background photons at Bob’s receiver. To investigate this trade-off, we have studied the effect of system parameters such as aperture size and FOV on the achievable distance and determined the optimal number of relay node (in the sense of maximizing the achievable distance). Our results show that the optimal number of relay increases as the FOV decreases and/or as the receive diameter decreases.

\balance
\bibliographystyle{IEEEtran}
\bibliography{reference}

\end{document}